\documentclass[12pt]{article}

\usepackage{epsf}
\usepackage{epsfig}
\usepackage{cite}
\usepackage{amsmath}
\usepackage{graphics}

\begin{document}

\setlength{\textheight}{21.5cm}
\setlength{\oddsidemargin}{0.cm}
\setlength{\evensidemargin}{0.cm}
\setlength{\topmargin}{0.cm}
\setlength{\footskip}{1cm}
\setlength{\arraycolsep}{2pt}

\renewcommand{\thefootnote}{\#\arabic{footnote}}
\setcounter{footnote}{0}

\newcommand{\gtrsim}{ \mathop{}_{\textstyle \sim}^{\textstyle >} }
\newcommand{\lesssim}{ \mathop{}_{\textstyle \sim}^{\textstyle <} }
\newcommand{\rem}[1]{{\bf #1}}
\renewcommand{\thefootnote}{\fnsymbol{footnote}}
\setcounter{footnote}{0}
\def\thefootnote{\fnsymbol{footnote}}

\hfill {\tt arXiv:0902.nnnn[astro-ph]}\\
\vskip .5in

\begin{center}

\bigskip
\bigskip

{\Large \bf Positron Excess, Luminous-Dark Matter Unification
and Family Structure}

\vskip .45in

{\bf Paul H. Frampton\footnote{frampton@physics.unc.edu}}

{\it Department of Physics and Astronomy, University of North Carolina,
Chapel Hill, NC 27599-3255.}

\vskip .30in

{\bf Pham Q. Hung\footnote{pqh@virginia.edu}}

{\it Department of Physics,
University of Virginia, Charlottesville, VA 22904-4714.}

\vskip .3in

\end{center}

\vskip .4in 
\begin{abstract}
It is commonly assumed that dark matter may be
composed of one or at most a few elementary particles.
PAMELA data present a window of opportunity
into a possible relationship between luminous and 
dark matter. Along with ATIC data the two positron
excesses are interpreted as a reflection of dark matter
family structure. In a unified model it
is predicted that at least a third enhancement
might show up at a different energy. The strength
of the enhancements however depends on interfamily mixing angles.
\end{abstract}

\renewcommand{\thepage}{\arabic{page}}
\setcounter{page}{1}
\renewcommand{\thefootnote}{\#\arabic{footnote}}

\newpage

\noindent {\it Introduction}

\bigskip

\noindent Since we are observing the energy density
of dark matter (DM) necessarily from the standpoint
of luminous matter, it may be worth imagining that
there exist large DM molecules which lead
to intelligent life including scientists who are
studying what we call luminous matter. Would
such hypothetical DM scientists be sufficiently Ptolemaic
to assume that luminous matter is made up
of only one or at most a few elementary particles
and what could
they possibly assume about our knowledge of
mathematics and biology?

\bigskip

\noindent Here we attempt a small step to
address this question which is potentially of
broad interest to mathematicians and biologists as
well as to theoretical physicists and observational
astrophysicists.

\bigskip

\noindent From observational data of galaxies and clusters
of galaxies it has been established that there exists
DM which interacts gravitationally \cite{silk}. There
is more DM than luminous matter energy density (a.k.a. baryonic matter)
by a factor of about five or six, i.e. $\Omega_{DM}/\Omega_{B} \sim 5-6$.

\bigskip

\noindent Because only gravitational interaction of DM
has been observed there is incompleteness
in understanding of the nature of the DM. Candidates
for its constituents have ranged in mass by many
orders of magnitudes all the way from a microelectronvolt
to significant fractions of a solar mass.

\bigskip

\noindent As always only observational data can provide
an arbiter and recent data from PAMELA\cite{PAMELA} and ATIC \cite{ATIC}
provide a window of opportunity on how
dark matter is unified with luminous matter
in a LUDUT (Luminous-Dark Unified Theory) containing 
a unification scale, with subscript truncated for convenience,
$M_{DUT}$. Taken at face value, the positron excess reported by PAMELA and
the lepton flux excess reported by ATIC appeared to occur at
two different energy scales. There might even be two
``bumps'' in the ATIC data-for a total of three
enhancements?- although the data are not sufficiently significant
at the present time.

\bigskip

\noindent The aforementioned excesses
could have at least three possible origins assuming
the data are correct. The excesses can arise from
(i) a nearby astrophysical object
such as a pulsar and involve no new physics;
(ii) annihilation of dark matter with dark antimatter;
or (iii) decay of dark matter.

\bigskip

\noindent In the present Letter we assume 
dark matter-antimatter annihilation is absent and that the positron
excess arises entirely from source (iii).

\bigskip

\noindent $M_{DUT}$ will be
generated by luminous-dark unification
involving the electroweak sector of the
standard model
with color $SU(3)$ regarded for simplicity
as a spectator. Already with this
simplification we see that the dark
matter must have a three-family structure
like the luminous matter\footnote{If there exists
a fourth luminous matter family \cite{FHS}
the same argument predicts a fourth dark matter family.}.
It is perhaps simpler to attribute the ``bumps''
in the PAMELA and ATIC data to two different
sources rather than one. If our interpretation is correct,
there could be at least one more enhancement at a different energy.
The amount of the excesses will depend, in our model, on
the mixing angles involved in the decay of the dark into luminous
matter.

\newpage

\noindent {\it Necessity for unification}

\bigskip

\noindent In order for dark matter to decay
into positrons there must be a non-gravitational
interaction between the relevant dark matter
particle $\chi$ and leptons. 

\bigskip

\noindent We denote the spin-1/2 dark matter field
by $\chi$ bound into a ($\bar{\chi} \chi)$ by
a strongly-coupled gauge interaction $SU(n)_{DM}$ 
similar to technicolor.
The annihilation
of $(\bar{\chi}\chi)$ into electron-positron takes place
by exchange of a gauge boson of $SU(n)_{DM}$
which we denote by $W_{DM}$.

\bigskip

\noindent It is immediate to see that $W_{DM}$ must
transform as a doublet under electroweak $SU(2)_L$
which already hints that unification of
$SU(n)_{DM}$ with $SU(2)_L$ underlies the non-gravitational
interaction of dark with luminous matter.

\bigskip

\noindent To accommodate the PAMELA\cite{PAMELA}
and ATIC\cite{ATIC} data requires that we study
the necessary lifetime $\tau_{\bar{\chi}\chi}$
for a dark-luminous unification scale $M_{DUT}$.

\bigskip

\noindent Assuming strong coupling $\alpha_{DM} = O(1)$ the
lifetime is
\begin{equation}
\tau_{\bar{\chi}\chi} \sim \left( \frac{M_{DUT}^4}{M_{\bar{\chi}\chi}^5} \right)
= \eta^4 (M_{\bar{\chi}\chi})^{-1}
\label{lifetime1}
\end{equation}
where $M_{DUT} = \eta M_{\bar{\chi}\chi}$. Given the data
it is natural assume $M_{\bar{\chi}\chi} \simeq 2$ TeV or 
$M_{\bar{\chi}\chi} \simeq 10^{-26}$ s and so
\begin{equation}
\tau_{\bar{\chi}\chi} \sim (10^{-26} s) \eta^4
\label{lifetime2}
\end{equation}

\bigskip

\noindent To estimate $\tau_{\bar{\chi}\chi}$ one can use the flux of $e^+$
reported by the observations\cite{PAMELA,ATIC}
to conclude that $\eta\sim10^{13}$\cite{Strumia,Arvanitaki}.

\newpage

\noindent {\it Unified Model}

\bigskip

\noindent The most economical way to unify the SM $SU(2)_L \times U(1)_Y$
with a general dark matter group $SU(n)_{DM}$ is by 
embedding these two groups as follows.
\begin{eqnarray}
\label{breaking}
SU(n+2) \times U(1)_S &\rightarrow& SU(n)_{DM} \times SU(2)_L \times 
U(1)_{DM} \times U(1)_S \nonumber \\
&\rightarrow& SU(n)_{DM} \times SU(2)_L \times 
U(1)_Y \nonumber \\
&\rightarrow& SU(n)_{DM} \times U(1)_{em}\,.
\end{eqnarray}
With color, the unifying group is $SU(3)_C \times SU(n+2) \times U(1)_S$.
For the first family the assignments
under 
$(SU(3)_C, SU(n+2), Y_S)$ are   
\begin{eqnarray}
&&(3, n+2, Y_{c,S})_L + (1, n+2, Y_S)_L \nonumber \\
&&+(3, n+2, Y_{c,S})_R + (1, n+2, Y_S)_R \nonumber \\
&&+ (3, 1, \left\{ \begin{array}{c} Y^u_S \\ Y^d_S \end{array} \right\})_L 
+ (3, 1, \left\{ \begin{array}{c} Y^u_S \\ Y^d_S \end{array} \right\})_R \nonumber \\
&&+(1, 1, Y_S)_L + (1, 1, Y_S)_R \nonumber \\
\label{assignments}
\end{eqnarray}
and for the second and third families this
pattern is repeated. This feature that the
dark matter has family structure
like the luminous matter will be
central in our interpretation of the
positron data from PAMELA and ATIC. 
Notice that the $SU(n+2)$ {\em nonsinglet} fermions
transform under the subgroup 
$SU(3)_C \times SU(n)_{DM} \times SU(2)_L \times U(1)_{DM} \times U(1)_S$
as 
\begin{eqnarray}
\label{nonsinglet}
&&(3,n,1,Y_{c,dm},Y_{c,S})_{L,R} + (3,1,2,Y_{q},Y_{c,S})_{L,R} \nonumber \\
&&(1,n,1,Y_{dm},Y_S)_{L,R} + (1,1,2,Y_{l},Y_S)_{L,R}
\end{eqnarray}

\bigskip

\noindent Three important remarks are in order concerning
the assignments in (\ref{assignments}).

\bigskip

\noindent First, the simplest
way to have an anomaly-free model in this case and to avoid
fermions with unconventional charges is to have both left and
right-handed fermions with identical transformations under the gauge group.

\bigskip

\noindent Second, if the dark matter strongly-coupled gauge group were to be
a QCD-like model, it is natural for it to be vector-like as it is
with $SU(3)_C$. 

\bigskip

\noindent Third, the above assignments involve right-handed
$SU(2)_L$ doublets for the ``luminous'' (non-dark) matter which
are absent in the SM. These ``mirror'' fermions have been used to
construct a model of electroweak-scale right-handed neutrinos with
wide implications for the seesaw mechanism at the LHC \cite{hung}.

\bigskip

\noindent In this work, our principal requirement is for dark matter
to be EW singlet. In view of the assignments in Eq.(\ref{nonsinglet}),
this translates into the requirement that the SM $U(1)_Y$ quantum
numbers of the dark matter particles should vanish. (The
color-nonsinglet dark matter particles quickly annihilate each
other with their density being drastically reduced by the Boltzmann
factor when the temperature drops below their masses (of order of a TeV).) 

\bigskip

\noindent To implement the above requirement, we now work out the generator
of $U(1)_{DM}$.
There are $n+1$ diagonal generators in $SU(n+2)$, one of which,
$\lambda_3 / 2$, can be taken to be $T_{3L}$ of $SU(2)_L$.
One combination of the remaining $n$ diagonal generators
$\lambda_8 / 2,..,\lambda_{n^2+4n+3}/ 2$ becomes the
$U(1)_{DM}$ generator. It is convenient to use the
following generators without the normalizing factors in front:
$T_8 = diag(1,1,-2,0,..,0); T_{15} = diag(1,1,1,-3,0,..,0);..;
T_{n^2+4n+3} = diag(1,1,...,-(n+1))$. Let us denote by
$T_{DM}$ the generator of $U(1)_{DM}$ which is given by
\begin{equation}
\label{TDM}
T_{DM} = \alpha_1 T_8 + \alpha_2 T_{15} +..+ \alpha_n T_{n^2+4n+3} \,.
\end{equation}
The hypercharge generator can most simply be written as
\begin{equation}
\label{hypercharge}
\frac{Y}{2} = y_{DM}\, T_{DM} + y_S I \,,
\end{equation}
where $I$ is the unit matrix and generator of $U(1)_S$. In order
for the dark matter to have vanishing hypercharge, i.e.
$Y/2 \propto  diag(1,1,0,..,0)$ where the first two elements (1)
refer to the SM doublet and the zeros refer to the DM, the coefficients
$\alpha_i$ in (\ref{TDM}) are found to be
\begin{equation}
\label{alpha}
\alpha_{n-k} = (\frac{y_S}{y_{DM}}) \frac{n+2}{(n-k+1)(n-k+2)} \,,
\end{equation}
with $k=0..n-1$. From (\ref{hypercharge}) and (\ref{alpha}), one can immediately
find the first two nonvanishing diagonal elements of $Y/2$, namely
$y_S \, \frac{n+2}{2}$ and consequently
\begin{equation}
\frac{Y}{2} = y_S ~~ diag( \frac{n+2}{2}, \frac{n+2}{2},0,..,0) .
\label{yS}
\end{equation}
From Eq.(\ref{yS}) one obtains the following conditions
on the $U(1)_S$ quantum numbers for the color singlets, $y_l$, and
for the color triplets, $y_q$:
\begin{eqnarray}
\label{condt1}
y_l \, \frac{n+2}{2}&=& -\frac{1}{2} \,, \nonumber \\
y_q \, \frac{n+2}{2}&=& \frac{1}{6} \,.
\end{eqnarray}
The electric charge is now $Q = T_{3L} + Y/2$ where $T_{3L}= \lambda_3 / 2$.

\bigskip

The next step is to determine the scale $M_{DUT}$ where the DM 
gauge group $SU(n)_{DM}$ becomes unified with the SM $SU(2)_L$. Let us
denote by $M_{DM}$ the scale where the $SU(n)_{DM}$ gauge group becomes strongly coupled.
At $M_{DUT}$, one has, by definition, $\alpha_{2}(M_{DUT}) = \alpha_{DM}(M_{DUT})
=\alpha_{DUT}$. The solution to the one-loop evolution equations gives
\begin{eqnarray}
\label{RG}
\alpha^{-1}_{2}(M_{DM}) &=& \alpha^{-1}_{DUT} + 8\,\pi\,b_2 \ln \frac{M_{DM}}{M_{DUT}} \,,\nonumber \\
\alpha^{-1}_{TC}(M_{DM}) &=& \alpha^{-1}_{DUT} + 8\,\pi\,b_{DM} \ln \frac{M_{DM}}{M_{DUT}}\,,
\end{eqnarray}
where $b_2 = (22-8\,n_G-n_{S,2})/48\pi^2$ and $b_{DM} = (11\,n-8\,n_G-n_{S,DM})/48\,\pi^2$ with
the factor of $8$ reflecting the fact that one has both left and right-handed fermions and
$n_G$ stands for the number of families. The scalar contributions are denoted by
the generic notations $n_{S,2}$ and $n_{S,DM}$
which do not reflect their group representations. The unification scale $M_{DUT}$
can be related to the DM scale $M_{DM}$ by
\begin{equation}
\label{MU}
M_{DUT} = M_{DM}\,\exp\{6\,\pi \, \frac{(\alpha^{-1}_{2}(M_{DM})-
\alpha^{-1}_{DM}(M_{DM}))}{(11\,n-22-(n_{S,DM}-n_{S,2}))}\} \,.
\end{equation} 
Notice that the expression in Eq.(\ref{MU}) is independent of the
number of families as expected. As mentioned above, the mass of the dark
matter particle is expected to be in the TeV range which will be taken
also to be the strongly-coupled DM scale. Hence $M_{DM} = O(TeV)$. 

\bigskip

\noindent Next,
in order to determine the appropriate $n$ and hence $SU(n)_{DM}$,
two more inputs are needed. We assume $\alpha_{DM}(M_{DM}) =1$. 
Extrapolating $\alpha_{2}^{-1}(M_Z) =29.44$ to $M_{DM} = O(TeV)$, one
obtains $\alpha_{2}^{-1}(1\,TeV) =29.19$ and $\alpha_{2}^{-1}(1\,TeV) =28.17$
for $n_G=3$ and $n_G=4$ respectively. 

\bigskip

\noindent Finally, we will require that
$M_{DM} < M_{DUT} < m_{Pl}$ with $M_{DUT} \sim 10^{15}-10^{16}$ GeV
in order to obtain a reasonable lifetime for the dark matter as discussed above.

\bigskip

\noindent
From (\ref{MU}) several results concerning a viable $SU(n)_{DM}$ are obtained.

\begin{itemize}

\item $n=2$ i.e. $SU(2)_{DM}$:

\noindent It is easy to see that this is an {\em unviable} scenario since one obtains either 
$M_{DUT} \gg m_{Pl}$ or $M_{DUT} \ll M_{DM}$. Conversely, one can see from Eq.(\ref{RG}) that
for three families $b_{DM} <0$ and, if $\alpha_{DUT} <1$ (i.e. perturbative), 
$\alpha_{DM}(M_{DM}) \ll \alpha_{DUT} <1$ implying that $SU(2)_{DM}$ is not confining at
$M_{DM} = O(TeV)$.

\item $n=3$ i.e. $SU(3)_{DM}$:

\noindent Again, one can deduce from (\ref{MU}) that $M_{DUT}
\gg m_{Pl}$. This is also not viable.

\item $n=4$ i.e. $SU(4)_{DM}$:

\noindent Let us denote the scalar contribution in 
Eq. (\ref{MU}) by $n_{S,DM}-n_{S,2} \equiv x$.
One obtains:
\begin{eqnarray}
\label{mu1}
M_{DUT}  &\approx& 2.6 \times 10^{13}\,GeV ; x=0 \nonumber \\
    &\approx& 5.4 \times 10^{15}\, GeV ; x=4 \nonumber \\
    &\approx& 3 \times 10^{16} \, GeV ; x=5
\end{eqnarray}
\noindent  A LUDUT based on $SU(3)_C \times SU(6)_{DUT} \times U(1)$ appears 
to be the best choice since, for $n \geq 5$, $M_{DUT}$ turns out
to be much smaller than the above values.

\end{itemize}

From hereon, we shall take the DM gauge group to be $SU(4)_{DM}$ and the
full gauge group at the scale $M_{DUT}$ is 
$SU(3)_C \times SU(6) \times U(1)_S$. At the scale $M_{DM}$, one has
$SU(4)_{DM} \times SU(3)_C \times SU(2)_L \times U(1)_Y$. The DM particles
denoted by $\chi$ above come in two kinds and under 
$SU(4)_{DM} \times SU(3)_C \times SU(2)_L \times U(1)_Y$ they transform as
\begin{eqnarray}
\label{chi}
\chi_l &=&(4,1,1,0)_{L,R} \,, \nonumber \\
\chi_q&=&(4,3,1,0)_{L,R} \,.
\end{eqnarray}

\bigskip

\noindent As we have mentioned above, the
$SU(4)_{DM}$ singlets come in two chiralities: the SM particles with
left-handed doublets and right-handed singlets, and the mirror fermions
which carry opposite chiralities. It is beyond the scope of the paper to
discuss the physical implications of the mirror fermions. Such discussions
can be found in \cite{hung}. We will concentrate instead on the DM particles.

\bigskip

\noindent When $SU(4)_{DM}$ becomes strongly coupled, there are two types of DM ``hadrons'':
those coming from the bound states of $\chi_l$ and those coming from the bound states
of $\chi_q$.

I) $\chi_l$ bound states:
\begin{itemize}

\item DM ``mesons'': $\bar{\chi}_l \,\chi_l$

\item DM ``baryons'': 4 $\chi_l$

\end{itemize}

II) $\chi_q$ bound states:
\begin{itemize}

\item DM ``mesons'': $\bar{\chi}_q \,\chi_q$

\item DM ``baryons'': 12 $\chi_q$ (both $SU(4)_{DM}$ and $SU(3)_C$ singlet).

\end{itemize}

\bigskip

\noindent All of these DM ``hadrons'' have masses in the TeV range. However, as the temperature
dropped below the color DM particles $\chi_q$, the Boltzman factor $\exp(-M_{DM}/T)$ drastically
reduced their number density since they were still in thermal equilibrium with the SM quarks
through QCD interactions. 

\bigskip

\noindent This leaves the DM ``hadrons'' coming from the color-singlet
$\chi_l$'s. We shall make some comments at the end of the paper concerning their relic
density. Let us for the moment examine what these DM ``hadrons'', which we will
denote by $M_{\chi_l}$ and $B_{\chi_l}$ from hereon, can do. However, since $B_{\chi_l}$
contains 4 $\chi_l$'s, it is twice as heavy as $M_{\chi_l}$ whose mass will be taken to
be of order $TeV$. For the positron excess consideration, only the decay of $M_{\chi_l}$
plays a role while in the minimal model considered here $B_{\chi_l}$ is stable.

\bigskip
\bigskip

\noindent {\it Decay of DM ``hadrons''}

\bigskip
\bigskip

The decays of the DM ``hadrons'' $M_{\chi_l}$ into the SM leptons and quarks
proceed through the couplings with the gauge bosons which belong to the
$SU(6)/(SU(2)_L \times SU(4)_{DM} \times U(1)_{DM})$ 
coset group and which we denote by $W_{DM}$.
These gauge bosons are assumed to have masses of the order of $M_{DUT}$. 
The relevant basic interactions
are $g_{DUT} W_{DM}^{2,\mu} \, \bar{l}\gamma_{\mu} \chi_l$ and $g_{DUT} W_{DM}^{1,\mu}\,
(\bar{\chi}_l \gamma_{\mu} \chi_l + \bar{l} \gamma_{\mu} l + \bar{q} \gamma_{\mu} q)$ 
where $W^{2,1}_{DM}$ with masses $M_{2,1}$
refers to gauge bosons which are 
doublets and singlets of $SU(2)_L$ respectively and where,
for notational purposes, chiralities are omitted. 

\bigskip

\noindent One obtains the following 4-fermion operators
\begin{equation}
\label{4fermi}
{\cal L}_{\chi\,f} = \frac{g_{DUT}^2}{M_{2}^2}(\bar{l}\gamma_{\mu} \chi_l )(\bar{\chi}_l \gamma^{\mu} l)
+ \frac{g_{DUT}^2}{M_{1}^2}(\bar{\chi}_l \gamma^{\mu} \chi_l)(\bar{l} \gamma_{\mu} l + \bar{q} \gamma_{\mu} q) \,.
\end{equation}
From (\ref{4fermi}), one expects the following decay modes
$M_{\chi_l} \rightarrow \bar{l}\,l, \bar{q}\,q$ 
with the leptonic modes being expected to dominate, in particular if $M_2 < M_1$.
One can estimate the lifetime of these DM ``hadrons'' to be approximately
\begin{equation}
\label{tau}
\tau \sim \alpha_{DUT}^{-2}\frac{M_{DUT}^4}{M_{DM}^5} \,.
\end{equation} 
The lifetime $\sim 10^{26}\,sec$ is obtained from the above formula with $\alpha_{DUT}^{-2}$
estimated to be $\sim 26$, $M_{DUT} \sim 10^{16}$ GeV and $M_{DM} = O(TeV)$.
Furthermore the DM ``mesons'' decay principally into $e^+ e^-, \mu^+ \mu^-, \tau^+ \tau^-$ since
the branching ratio into quarks is expected to be smaller. 
This could possibly fit
into the observation that the $\bar{p}$ flux is in agreement with background \cite{PAMELA}

One important remark is in order at this point. For simplicity, mixing angles
between different families are omitted in (\ref{4fermi}). However, it goes
without saying that DM of different families decay into luminous matter with
different strengths because of these mixing angles. As a result, the
enhancements at different energies (i.e. different DM masses) are not expected
to be uniform: there might well be more enhancement at one energy as compared with
another.

\bigskip
\bigskip
\bigskip

\noindent {\it Summary and Discussion}

\bigskip

\noindent The most striking prediction of our model comes 
from the family structure of the DM which follows inevitably 
from the LUDUT theory unifying  at a large scale $M_{DUT}$. 

\bigskip

\noindent We predict excesses
at three mass locations (for three families of DM): 
one at the PAMELA peak, another at the (yet-to-be
confirmed) ATIC peak and a third one which is yet to be 
observed and which is expected to be
located at a mass not too far from the PAMELA and ATIC peaks.
The amount of enhancement is expected to vary
for different energies because of interfamily mixing angles.
In our model, this might be the reason why the excess at PAMELA
is larger than that at ATIC. There might already be a third
excess between the two ``bumps'' of PAMELA and ATIC, albeit
a supressed one because of small mixing angles; it
is not yet statistically significant.

\bigskip

\noindent

\noindent Such observations could provide a smoking gun for
a LUDUT
theory which unifies luminous and dark matter and
greatly expedite further understanding of the dark
sector.

\bigskip

\noindent Specific candidates for DM particles
have appeared in model building \cite{conformality}; also,
part of the DM energy density can arise
from intermediate mass black holes \cite{IMBH}. Other
work on the observed positron excess, none of which mentions
family structure, is exemplified in \cite{other}.

\bigskip

\noindent Last but not least, our model contains DM particles
which carry color but whose number density is greatly suppressed
by the Boltzman factor as we have mentioned above. However, these
particles can be produced at the LHC through the gluon fusion process,
albeit with a small cross section due to the high mass (TeV's). 
The signatures of these particles are under investigation.

\bigskip

\noindent More generally we believe that there is every reason
that Nature be equally as imaginative in the dark
side of the Universe as the luminous side. Thus,
while our DM family structure may appear as speculation,
the truth about DM could be even more complicated.
Pauli said after reading one paper
''these ideas are crazy but are they crazy enough?" 
and the remark may be germane also to the present one.

\begin{center}

\section*{Acknowledgements}

\end{center}

PHF was supported in part 
by the U.S. Department of Energy under Grant
No. DE-FG02-06ER41418 and PQH was supported in part by
the U.S. Department of Energy under Grant No. DE-FG02-97ER41027.

\bigskip
\bigskip
\bigskip
\bigskip


\bigskip
\bigskip
\bigskip
\bigskip
\bigskip
\bigskip
\bigskip
\bigskip
\bigskip


\begin{thebibliography}{99}

\bibitem{silk}
 For a review see G.~Bertone, D.~Hooper and J.~Silk,
  Phys.\ Rept.\  {\bf 405}, 279 (2005)
  [arXiv:hep-ph/0404175].
\bibitem{PAMELA}
PAMELA Collaboration, {\tt arXiv:0810.4995}.

\bibitem{ATIC}
ATIC Collaboration, Nature {\bf 456,} 362 (2008).

\bibitem{FHS}
For a review, see P.H. Frampton, P.Q. Hung and M. Sher, Phys. Reports {\bf 330,} 263 (2000).
{\tt hep-ph/9903387}. For recent constraints, see e.g.
  G.~D.~Kribs, T.~Plehn, M.~Spannowsky and T.~M.~P.~Tait,
  Phys.\ Rev.\  D {\bf 76}, 075016 (2007)
  [arXiv:0706.3718 [hep-ph]];
  P.~Q.~Hung and M.~Sher,
  Phys.\ Rev.\  D {\bf 77}, 037302 (2008)
  [arXiv:0711.4353 [hep-ph]];
  M.~Bobrowski, A.~Lenz, J.~Riedl and J.~Rohrwild,
  arXiv:0902.4883 [hep-ph].
Earlier works on the 4th generation include
  W.~S.~Hou, R.~S.~Willey and A.~Soni,
  Phys.\ Rev.\ Lett.\  {\bf 58}, 1608 (1987)
  [Erratum-ibid.\  {\bf 60}, 2337 (1988)];
  W.~S.~Hou, A.~Soni and H.~Steger,
  Phys.\ Rev.\ Lett.\  {\bf 59}, 1521 (1987).
\bibitem{Strumia}
E. Nardi, F. Sannino and A. Strumia, JCAP 0901:043 (2009).
{\tt arXiv:0811.4153 [hep-ph]}.

\bibitem{Arvanitaki}
A. Arvanitaki, S. Dimopoulos, S. Dubovsky,
P.W. Graham, R. Harnik and S. Rajendran,
{\tt arXiv:0812.2075 [hep-ph]}.

\bibitem{hung}
P.Q. Hung, Phys. Lett. {\bf B649}, 275 (2007). {\tt hep-ph/0612004};\\
Phys. Lett.  {\bf B659,} 585 (2008). {\tt arXiv:0711.0733 [hep-ph]};\\
Nucl. Phys. {\bf B805,} 326 (2008). {\tt arXiv:0805.3486 [hep-ph]}.

\bibitem{conformality}
P.H. Frampton, Mod. Phys. Lett. {\bf A22,} 931 (2007).
{\tt astro-ph/0607391};
  P.~Q.~Hung,
  Nucl.\ Phys.\  B {\bf 747}, 55 (2006)
  [arXiv:hep-ph/0512282];
  M.~Adibzadeh and P.~Q.~Hung,
  Nucl.\ Phys.\  B {\bf 804}, 223 (2008)
  [arXiv:0801.4895 [astro-ph]].

\bibitem{IMBH}
P.H. Frampton, {\tt arXiv:0806.1707[gr-qc]};\\
{\tt arXiv:0903.0113[astro-ph]}.

\bibitem{other}
V. Barger {\it et al.} Phys. Lett. {\bf B672,} 141 (2009).\\
{\tt arXiv:0809.0162[hep-ph]}.\\
N.Arkani-Hamed {\it et al.} Phys. Rev. {\bf D79,} 015014 (2009).\\
{\tt arXiv:0810.0713[hep-ph]}.\\
D. Feldman, Z. Liu and P. Nath. {\tt arXiv:0810.5762[hep-ph]}.
\\
E. Ponton and L. Randall. {\tt arXiv:0811.1029[hep-ph]}.\\
L. Bergstrom, {\it et al.} {\tt arXiv:0812.3895[astro-ph]}.\\
M. Ibe, Y. Nakayama, H. Murayama and T. Yanagida.
{\tt arXiv:0902.2914[hep-ph]}.


\end{thebibliography}
\end{document}